\documentclass[conference]{IEEEtran}
\IEEEoverridecommandlockouts
\usepackage{cite}
\usepackage{amsmath,amssymb,amsfonts}
\usepackage{algorithmic}
\usepackage{graphicx}
\usepackage{textcomp}
\usepackage{xcolor}
\usepackage{url}
\usepackage{comment}
\usepackage{float}
\usepackage{lipsum}  
\usepackage{lastpage}
\usepackage{multirow}
\usepackage{graphicx}
\usepackage{subcaption}
\usepackage{caption}
\usepackage[colorlinks=true, urlcolor=blue, linkcolor=red, citecolor=blue]{hyperref}
\def\BibTeX{{\rm B\kern-.05em{\sc i\kern-.025em b}\kern-.08em
    T\kern-.1667em\lower.7ex\hbox{E}\kern-.125emX}}
\begin{document}

\title{Developing a Cost-Effective Spectrometer: \\A Practical Approach}

\author{
\IEEEauthorblockN{
Zafrin Jahan Nikita\textsuperscript{1},
Mohammad Tahsin Alam\textsuperscript{1},
Yasir Mahmud\textsuperscript{1},
Rafichha Yasmin\textsuperscript{1},\\
Fairuz Darimi Bushra\textsuperscript{1},
Sheikh Hasib Cheragee\textsuperscript{1},
and Redwanul Talukder Zeem\textsuperscript{1}
}
\IEEEauthorblockA{
\textsuperscript{1}\textit{Department of Electrical and Electronic Engineering,} \\
\textit{Bangladesh University of Engineering and Technology (BUET), Dhaka 1205, Bangladesh} \\
\textsuperscript{*}Corresponding author: tahsin0799@gmail.com
}

}

\maketitle

\begin{abstract}

The paper demonstrates the design and execution of a low-cost optical spectrometer that employs a webcam, diffraction grating \& Python(a free, open-source programming language). The device's design prioritized economy and usability, with a black box casing to reduce stray light and increase measurement accuracy. A diffraction grating made from a DVD was used to split light into its constituent wavelengths, which were then collected by the camera. The calibration procedure used a RED TIDE USB650 Fiber Optic Spectrometer to set calibration value for various wavelength ranges, which ensured that the spectrometer's results closely matched those derived from the former, a high-cost industry-standard model. Spectrums of several light sources, such as red, green, blue, yellow, white, magenta, orange and UV LEDs, as well as a green laser, were studied and compared. The results showed a high level of precision, with minimal divergence from industry-standard spectrometer measurements. This comparison was carried out utilizing Origin software, which allowed for extensive analysis and display of the spectrum data. In addition, the spectrometer captured data in real time, allowing users to watch live spectrum changes and ensure instant accessibility of results. Despite its simplicity and low cost, the spectrometer provides significant value for instructional and practical applications, making it a valuable tool in cost-constrained situations.

\end{abstract}


\section{Introduction} Spectrometry, the measurement of light intensity across wavelengths, is essential in fields like chemistry, physics, biology, and environmental science. Traditional spectrometers, while accurate, are often expensive, limiting accessibility for education and low-budget research. Developing affordable alternatives is crucial to democratizing this technology.

Several studies have attempted low-cost spectrometer designs with varying success. Silva et al. \cite{daSilva} created a portable spectrometer using a CMOS image sensor, a high-intensity LED, and transmission grating, processed via OpenCV on a Raspberry Pi. While innovative, this design faced challenges in accuracy, resolution, limited wavelength range (380-700 nm), sensitivity to sample impurities, and calibration issues \cite{hong}.

Likith et al. \cite{likith} developed an even more economical spectrometer using cardboard, DVDs, and webcams. However, this design suffered from poor accuracy and resolution, durability issues, and alignment difficulties for non-experts. The webcam's infrared filter also limited IR detection, and converting images to greyscale distorted spectral intensities, reducing accuracy \cite{fieldSpectrometerHomeMaterials2020} \cite{technovationDIYLowCost}.

In the work by Barzegar et al. \cite{barzegar2019}, a smartphone-based spectrometer was developed using a 3D-printed holder and diffraction grating film, which provided satisfactory performance for educational purposes but was limited in wavelength range and spectral resolution. Zhang et al. \cite{zhang2018} proposed a smartphone-based system utilizing a micro-electro-mechanical system (MEMS) grating, achieving compactness and portability but requiring expensive fabrication processes.

Other approaches include the work by Neira et al. \cite{neira2020}, where a fiber-optic based spectrometer was created for environmental monitoring. Although offering good resolution, the need for precise fiber alignment and the high cost of fiber components limited its application in low-budget scenarios. Wei et al. \cite{wei2019} developed a UV-visible spectrometer for chemical analysis using low-cost materials, but the device's performance degraded significantly outside the visible range due to the limitations of the sensor used.

Further developments by Martínez-Pérez et al. \cite{martinez2019} utilized a custom-designed slit and transmission grating on a CCD-based detector, achieving good spectral range and resolution but increasing complexity and cost. Similarly, Moudgil et al. \cite{moudgil2018} demonstrated a compact spectrometer for bio-sensing applications using a holographic grating, but the high cost of holographic components posed a significant barrier.

More recently, Singh et al. \cite{singh2021} proposed a modular spectrometer design using a Raspberry Pi and a linear CCD array, which enhanced flexibility and resolution but required advanced programming skills for operation and calibration. The work of Ghafoor et al. \cite{ghafoor2020} also showed promise in developing a low-cost spectrometer for agricultural applications, but faced challenges with calibration and the influence of environmental factors on the results.

Building upon these foundations, this paper presents an improved design and execution of a low-cost optical spectrometer that addresses many of the limitations noted in previous studies. Our approach uses a webcam, diffraction grating, and Python—a free, open-source programming language. The primary focus of this design is to balance economy with usability, ensuring that the resulting device is both affordable and functional.

Key improvements in our design include enhanced accuracy and resolution by using a DVD as a diffraction grating and a well-calibrated RED TIDE USB650 Fiber Optic Spectrometer. This ensures higher precision compared to the models by Silva et al. \cite{daSilva} and Likith et al. \cite{likith}. The spectrometer's durability and stability are improved with a robust black box casing, which reduces stray light and enhances measurement accuracy. Additionally, our device captures a broader wavelength range, addressing the limitations in Silva et al.'s design. The simplified alignment process and straightforward calibration make it accessible to non-experts.

To validate performance, spectra from various light sources—including different colored LEDs and a green laser—are captured and analyzed. These are compared to results from a high-cost, industry-standard spectrometer using Origin software for analysis. Our findings show high precision with minimal divergence, highlighting the device’s potential for practical use. Real-time data capture also enables users to observe live spectrum changes.

This improved design addresses the limitations of earlier models and offers significant value for instructional and practical applications, particularly in cost-constrained settings.

\section{Key Components}

\begin{itemize}
    \item Housing module(black box)
    \item Diffraction grating
    \item Camera
    \item Blades(as slit apertures)
    \item Python software
\end{itemize}

\begin{figure}[htbp]
    \centering    \includegraphics[width=\linewidth]{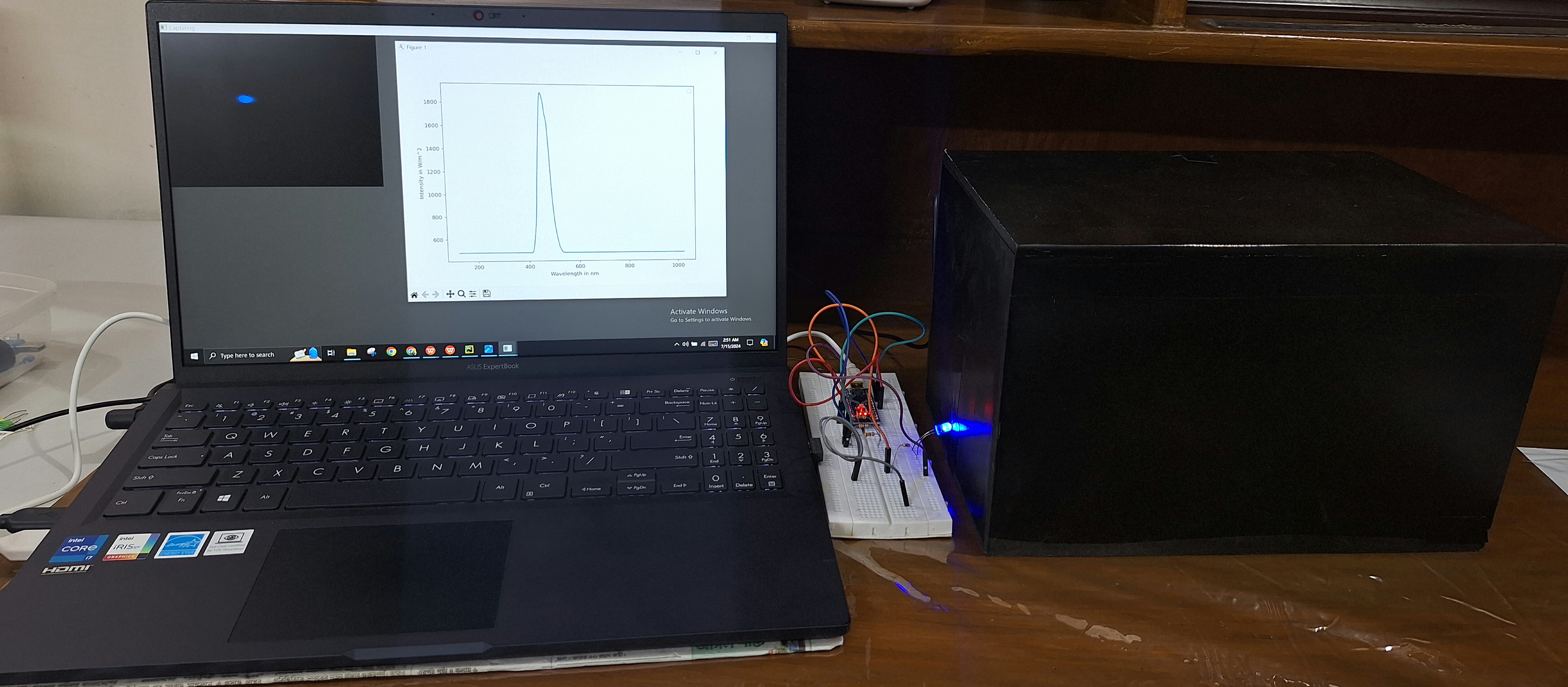}
    \caption{Overall Setup}
   \label{Experimental Setup}
\end{figure}

\begin{figure}[htbp]
    \centering
    \includegraphics[width=0.9\linewidth]{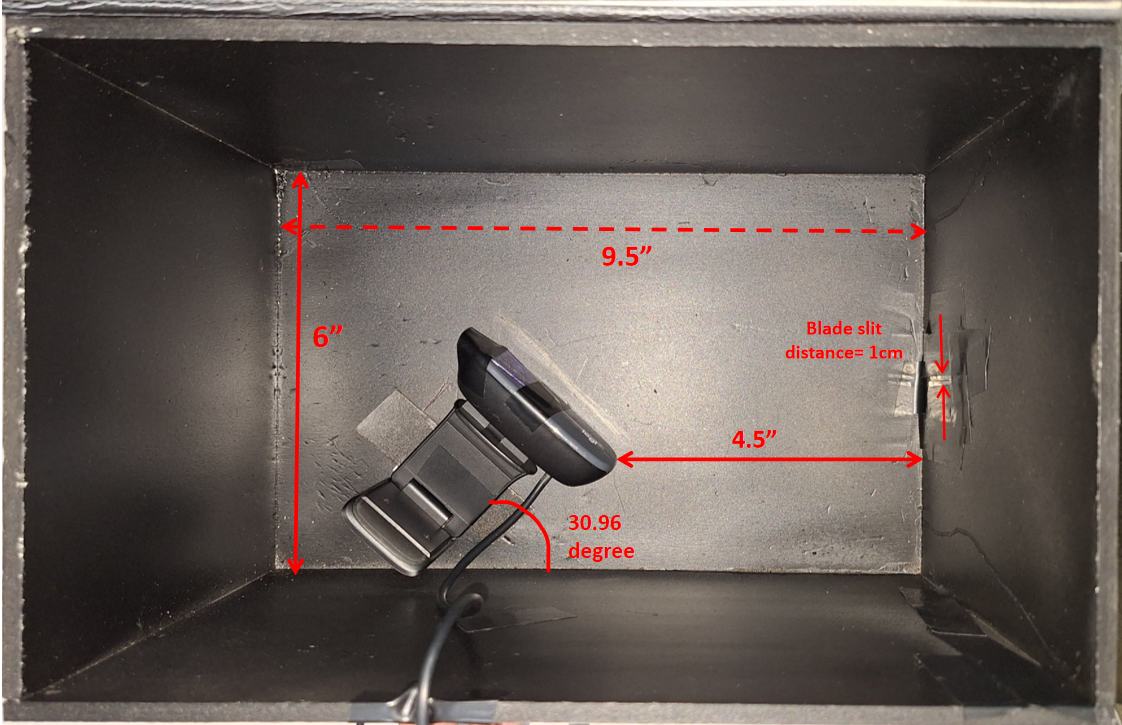}
    \caption{Dimensions inside housing}
    \label{inside_dim}
\end{figure}

\begin{figure}[htbp]
    \centering
    \includegraphics[width=0.5\linewidth, trim=5 5 10 10, clip]{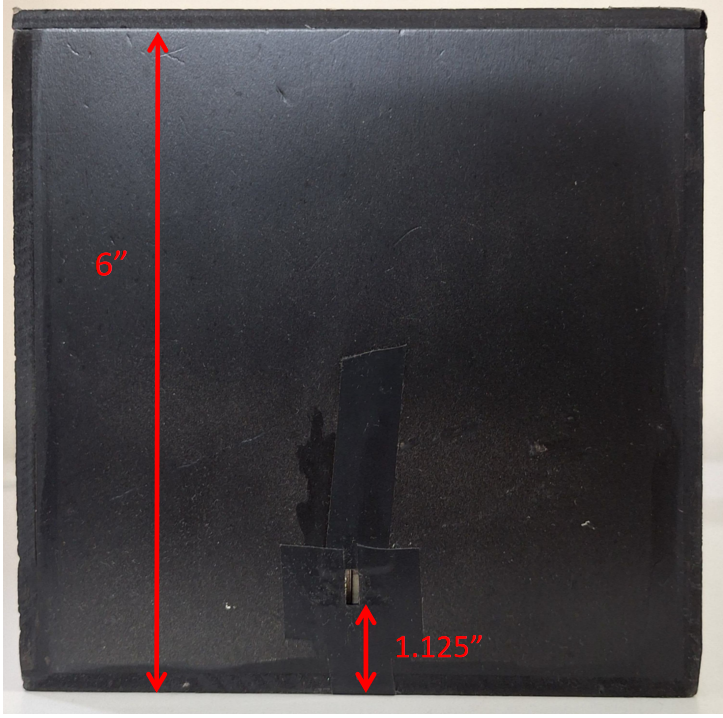}
    \caption{Dimensions outside box}
    \label{outside_dim}
\end{figure}

\section{Experimental Setup \& Workflow}

\subsection{\textbf{Hardware setup}}

Figure \ref{Experimental Setup} represents the Housing (black box) as well as the output screen of Python program.The output screen allows the user to view the color captured on the grating by the webcam as well as the output spectrum in a live stream.Dimensions of the hardware setup is shown in Figure \ref{inside_dim} and Figure \ref{outside_dim}.\\ We used a black box for the experiment to achieve stray light reduction, improved sensitivity, controlled environment \& modular design.The black box absorbs stray light, ensuring accurate readings by creating a controlled environment for the intended optical path. Minimizing stray light enhances the signal-to-noise ratio, crucial for detecting weak signals and analyzing faint light sources.It also provides a stable environment, maintaining calibration and minimizing drifts due to temperature changes or ambient light fluctuations.Using pre-built modules or kits in black boxes simplifies construction and reduces the need for complex mechanical or optical skills.

\subsection{\textbf{Process steps}}

The light emitted from an LED or laser is focused into a narrow beam and directed through a 1 cm slit into a black housing. A webcam, positioned at a 30.96-degree angle relative to the wall as shown in Figure \ref{inside_dim}, captures the light. To isolate the first-order diffraction pattern, a diffraction grating is placed in front of the camera lens. This grating disperses the light into its component wavelengths by allowing only certain wavelengths to constructively interfere in a specific direction, while others are canceled out due to destructive interference. DVDs, which contain a built-in diffraction grating beneath their silver layer, are used for this purpose. The grating is affixed to the camera, which is housed to ensure accurate capture of the diffraction pattern.

\subsection{\textbf{Calibration}}

In our experiment, we used white, red, and green LEDs to measure their spectra using the \textbf{RED TIDE USB650 Fiber Optic Spectrometer}. The goal was to calibrate our device accurately.

At the beginning of the program, the initial calibration value was set to zero (\texttt{calib} = 0). After running the experiment for the first time, we detected the peak emission wavelengths for each LED. Based on the emission data, we adjusted the calibration values by adding or subtracting appropriate values, verified using the RED TIDE USB650 Fiber Optic Spectrometer.

\begin{figure*}[htbp]
    \centering
    \begin{subfigure}{0.32\textwidth}
        \includegraphics[width=\linewidth]{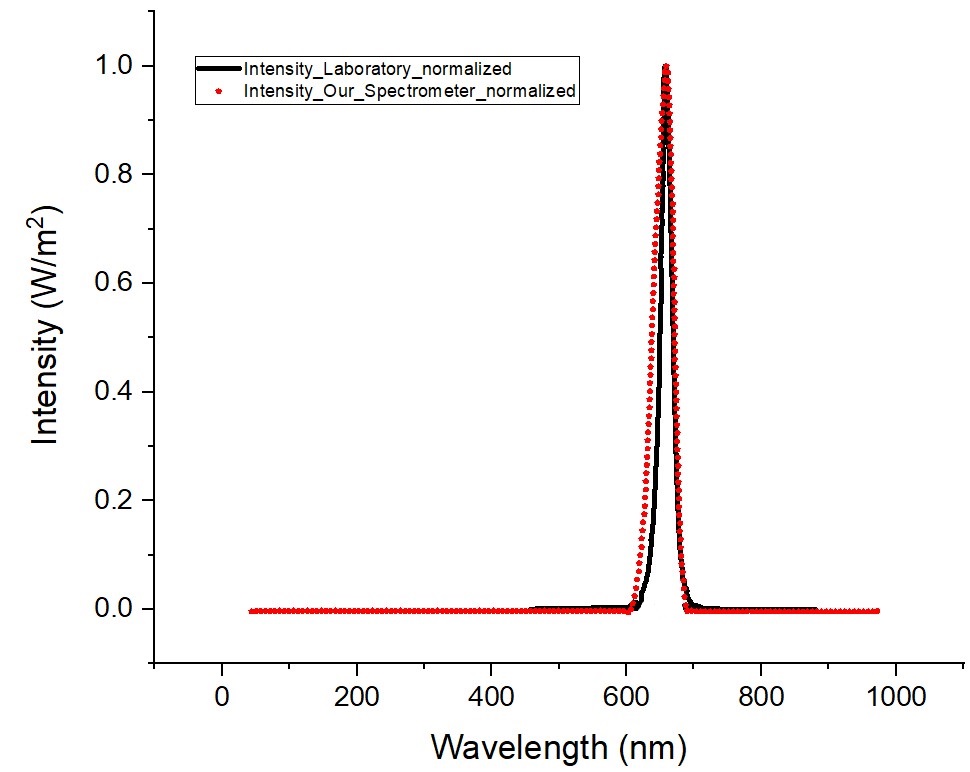}
        \caption{Red LED}
        \label{RedLEDcomparison}
    \end{subfigure}
    \begin{subfigure}{0.32\textwidth}
        \includegraphics[width=\linewidth]{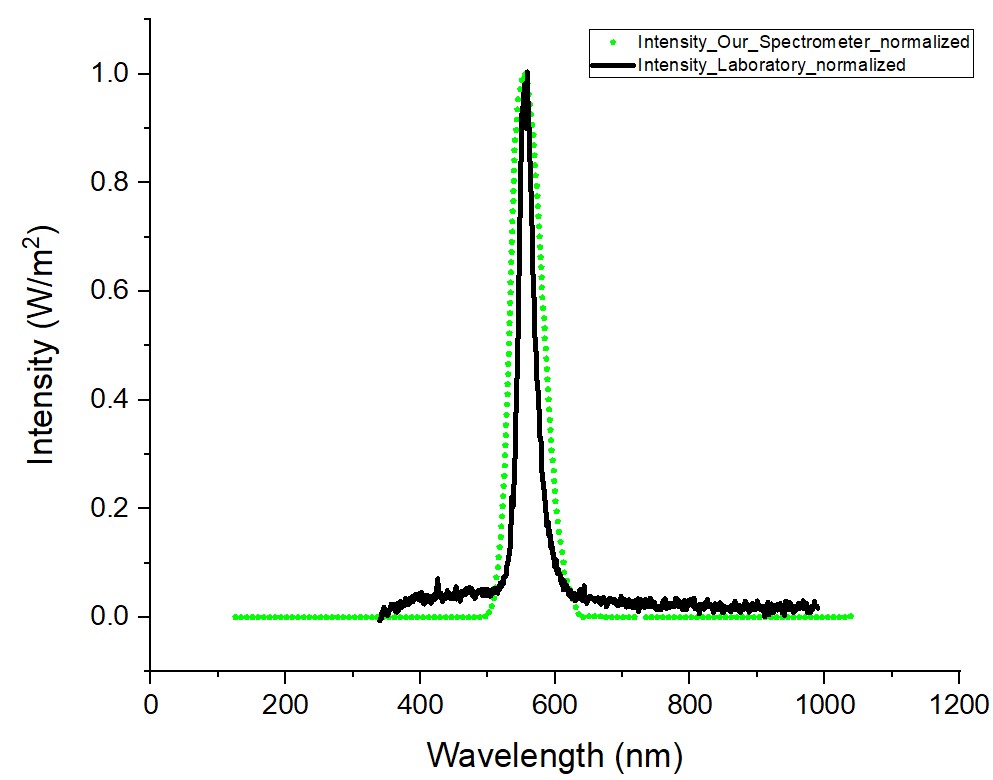}
        \caption{Green LED}
        \label{GreenLEDcomparison}
    \end{subfigure}
    \begin{subfigure}{0.32\textwidth}
        \includegraphics[width=\linewidth]{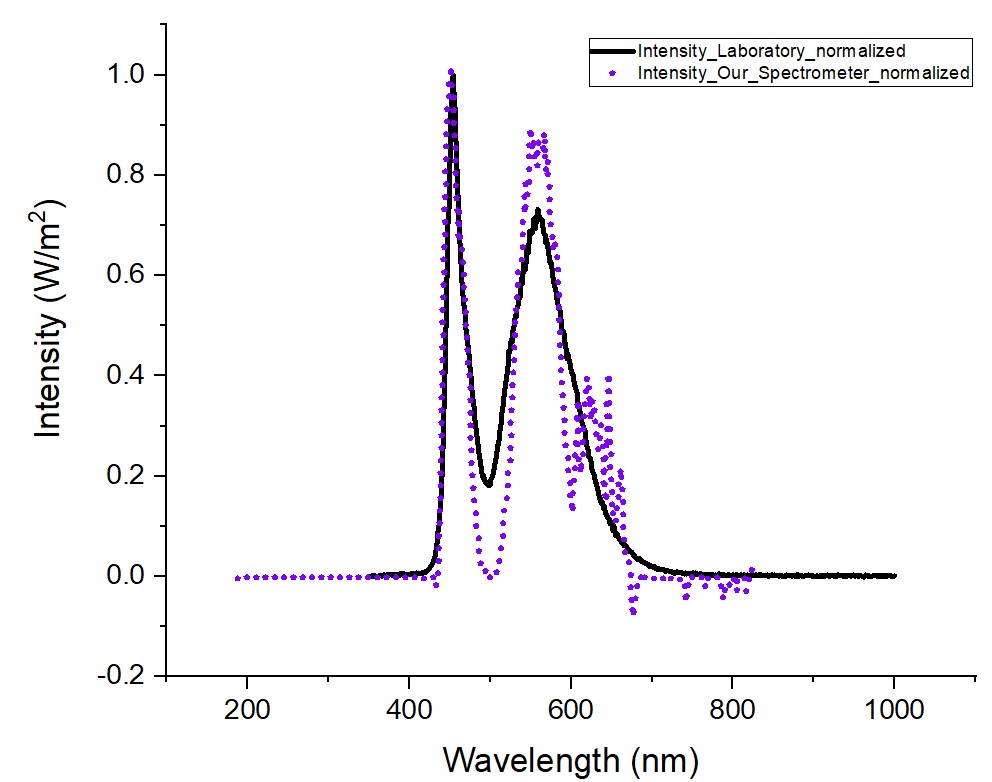}
        \caption{White LED}
        \label{WhiteLEDcomparison}
    \end{subfigure}
    \caption{Spectral comparison of selected LEDs with standard measurements.}
    \label{LEDcomparisons}
\end{figure*}

For the red LED, which operates in the wavelength range $600 \, \text{nm} < \lambda < 1200 \, \text{nm}$, a calibration value of \texttt{calib} = -35 produced the best outcome. This result is illustrated later in Figure \ref{RedLEDcomparison}.

As for the green and yellow LEDs, which operate in the range $480 \, \text{nm} < \lambda < 600 \, \text{nm}$, no calibration was necessary. We set \texttt{calib} = 0 in the decision loop as the results matched our expectations.

For the green laser (Model: \textbf{Geepass Laser 303}), the output emission wavelength was fully satisfactory and matched its rated specification.

Finally, for the blue and UV regions, corresponding to $300 \, \text{nm} < \lambda < 480 \, \text{nm}$, a calibration value of \texttt{calib} = 27 provided the expected results.

These calibration steps ensured the accurate performance of our spectrometer across the various wavelength ranges.

\begin{table}[]

\caption{Experimental Data}
\label{Experimental Data}
\renewcommand{\arraystretch}{1.6}

\begin{tabular}{|c|c|c|c|}
\hline
{ \textbf{Source Type}}                                & {\textbf{\begin{tabular}[c]{@{}c@{}}Peak Wavelength(s) \\ (nm)\end{tabular}}} & {\textbf{Line width(nm)}}                                   \\ \hline
Red LED                                                                    & 658                                                                                                 & 671.5-637.7= 33.8                                                                  \\ \hline
Green LED                                                                  & 553                                                                                                 & 584.62-531.5=53.12                                                                 \\ \hline
Blue LED                                                                   & 462                                                                                                 & 498.5-441.9=56.6                                                                   \\ \hline
Orange LED                                                                 & 597.5                                                                                               & 611.5-586.2=25.3                                                                   \\ \hline
Yellow LED                                                                 & 502 \& 660                                                                                          & \begin{tabular}[c]{@{}c@{}}537.02-459.61=77.41;\\ 681.75-624.53=57.22\end{tabular} \\ \hline
White LED                                                                  & 451 \& 558                                                                                          & \begin{tabular}[c]{@{}c@{}}474.15-438.4=35.75 ;\\ 601.85-528.88=72.97\end{tabular} \\ \hline
\begin{tabular}[c]{@{}c@{}}Green Laser\\ (rating:532 ± 10 nm)\end{tabular} & 541                                                                                                 & 539-542=3                                                                          \\ \hline
UV LED                                                                     & 354                                                                                                 & 360.8-345=15.8                                                                     \\ \hline
Magenta LED                                                                & 407 \& 660                                                                                          & \begin{tabular}[c]{@{}c@{}}435-392=43;\\ 680-625=55\end{tabular}                   \\ \hline
\end{tabular}
\end{table}

\begin{figure*}[htbp]
    \centering
    \begin{subfigure}{0.32\textwidth}
        \includegraphics[width=\linewidth]{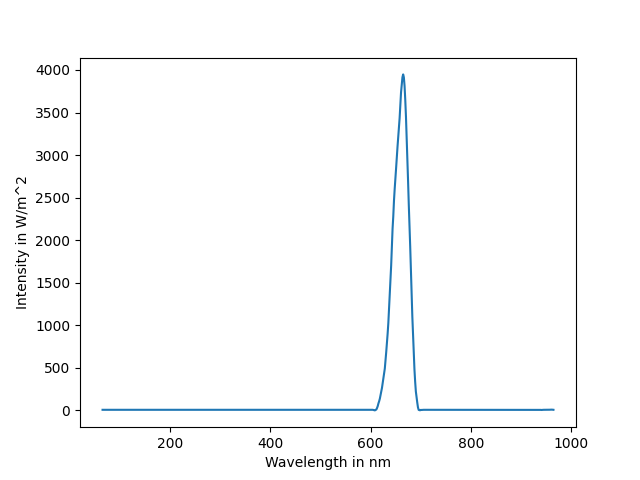}
        \caption{Red LED}
        \label{fig:redLED}
    \end{subfigure}
    \begin{subfigure}{0.32\textwidth}
        \includegraphics[width=\linewidth]{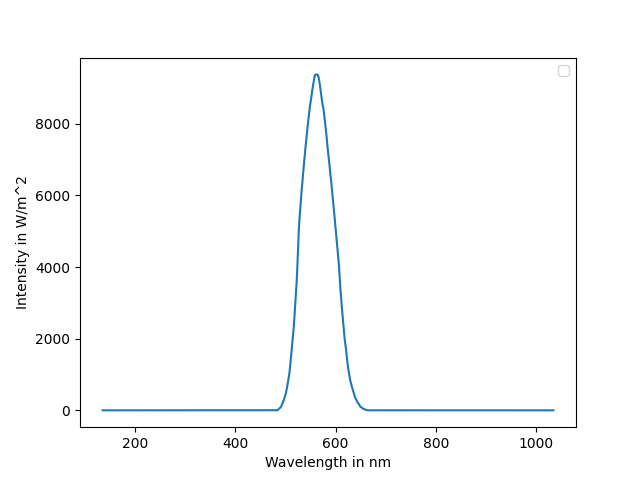}
        \caption{Green LED}
        \label{fig:greenLED}
    \end{subfigure}
    \begin{subfigure}{0.32\textwidth}
        \includegraphics[width=\linewidth]{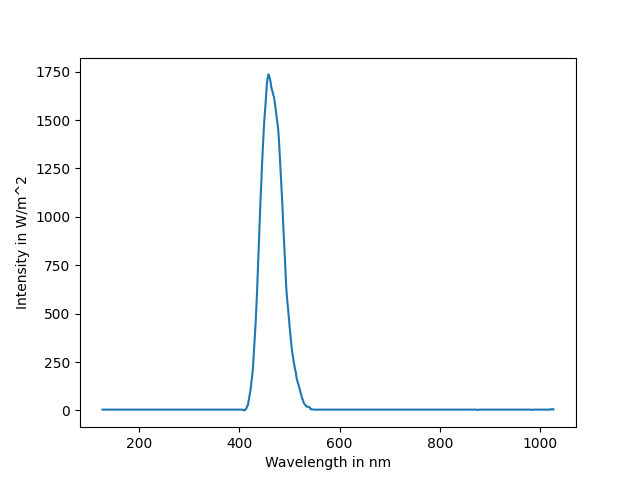}
        \caption{Blue LED}
        \label{fig:blueLED}
    \end{subfigure}
    
    \vspace{2mm}
    
    \begin{subfigure}{0.32\textwidth}
        \includegraphics[width=\linewidth]{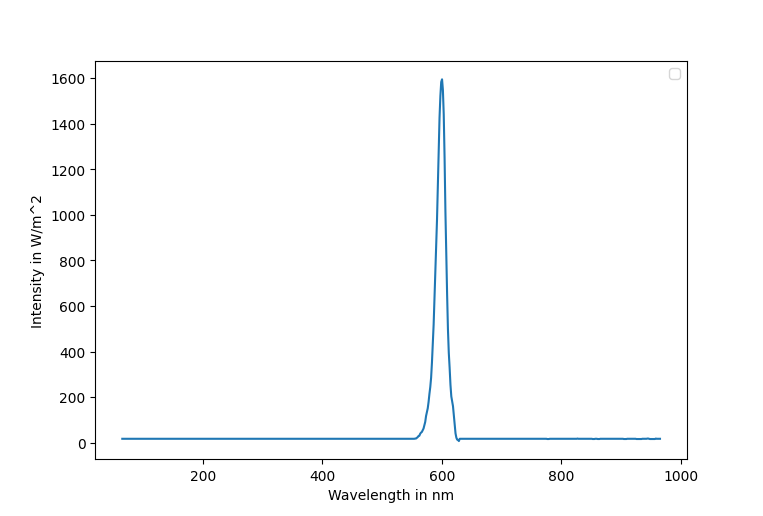}
        \caption{Orange LED}
        \label{fig:orangeLED}
    \end{subfigure}
    \begin{subfigure}{0.32\textwidth}
        \includegraphics[width=\linewidth]{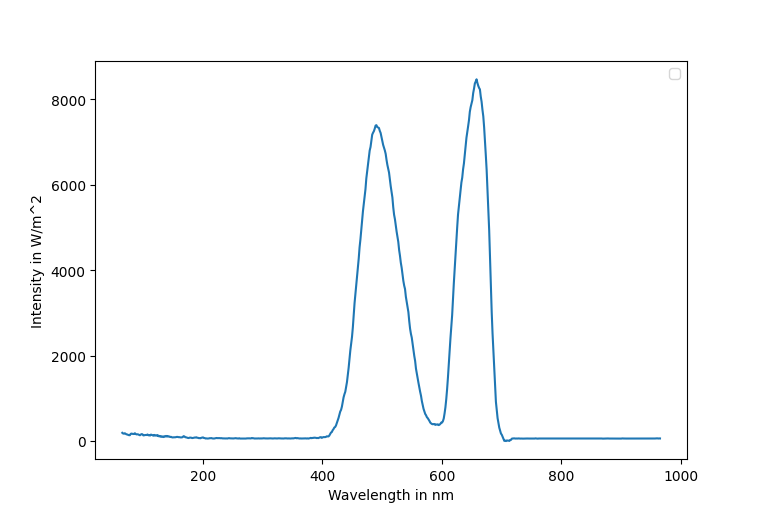}
        \caption{Yellow LED}
        \label{fig:yellowLED}
    \end{subfigure}
    \begin{subfigure}{0.32\textwidth}
        \includegraphics[width=\linewidth]{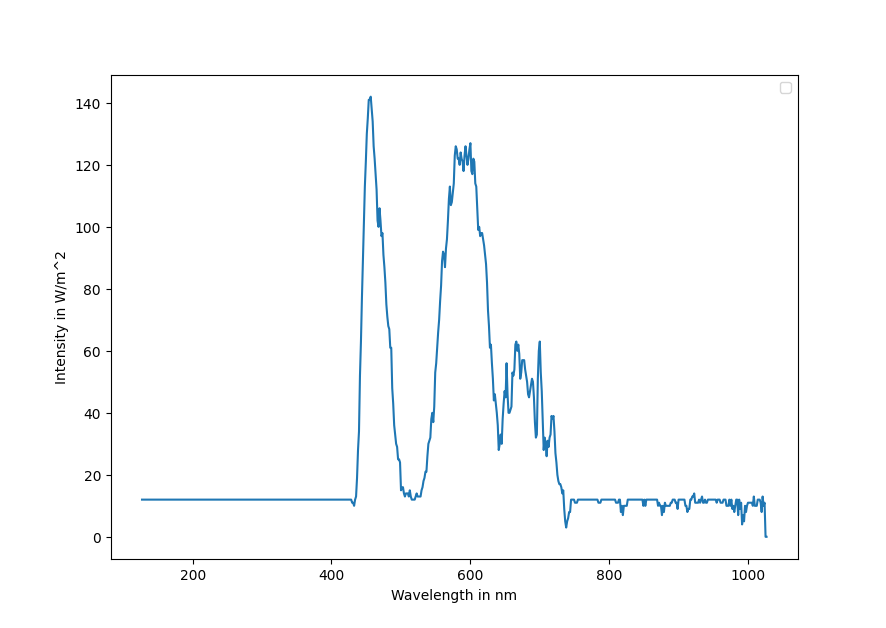}
        \caption{White LED}
        \label{fig:whiteLED}
    \end{subfigure}
    
    \vspace{2mm}
    
    \begin{subfigure}{0.32\textwidth}
        \includegraphics[width=\linewidth]{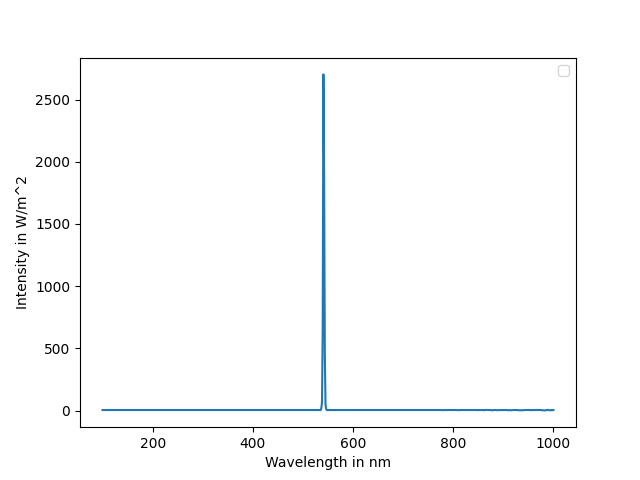}
        \caption{Green laser}
        \label{fig:greenLaser}
    \end{subfigure}
    \begin{subfigure}{0.32\textwidth}
        \includegraphics[width=\linewidth]{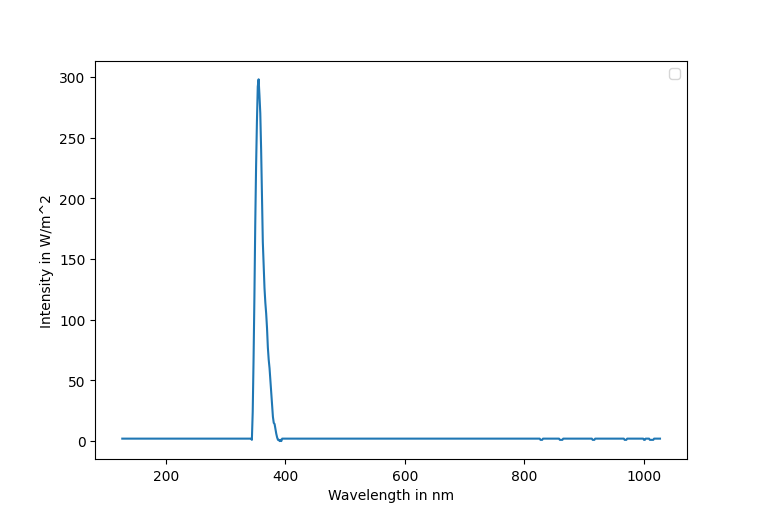}
        \caption{UV LED}
        \label{fig:UVLED}
    \end{subfigure}
    \begin{subfigure}{0.32\textwidth}
        \includegraphics[width=\linewidth]{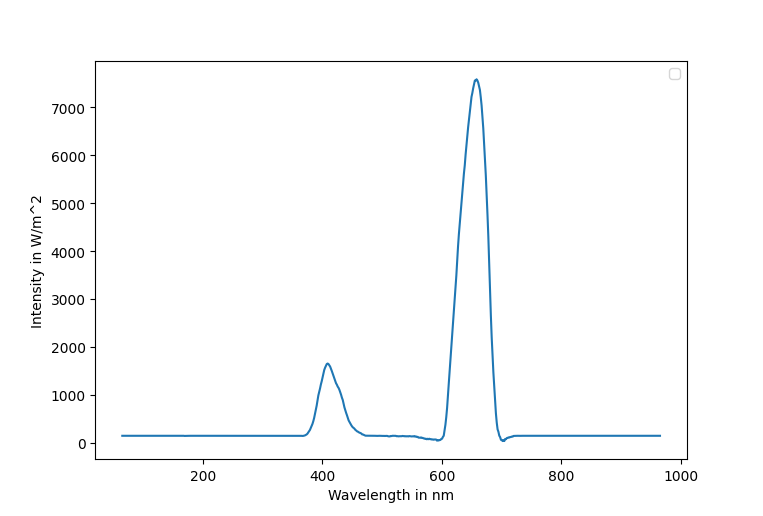}
        \caption{Magenta LED}
        \label{fig:magentaLED}
    \end{subfigure}
    
    \caption{Spectra of various light sources including LEDs and laser.}
    \label{allSpectra}
\end{figure*}

\begin{table*}[htbp]

\centering

\caption{Comparison with Red Tide USB650 Fiber Optic Spectrometer}
\label{Comparison table}
\renewcommand{\arraystretch}{1.5}

\begin{tabular}{|c|c|c|c|c|c|c|}
\hline
\textbf{Source type} & \textbf{\begin{tabular}[c]{@{}c@{}}Peak wavelength \\ (Our spectrometer)\end{tabular}} & \textbf{\begin{tabular}[c]{@{}c@{}}Peak wavelength\\  (Laboratory)\end{tabular}} & \textbf{\begin{tabular}[c]{@{}c@{}}Error in \\ peak wavelength\end{tabular}} & \textbf{\begin{tabular}[c]{@{}c@{}}Line width \\ (Our spectrometer)\end{tabular}}             & \textbf{\begin{tabular}[c]{@{}c@{}}Line width \\ (Laboratory)\end{tabular}} & \textbf{\begin{tabular}[c]{@{}c@{}}Error in \\ line width\end{tabular}}     \\ \hline
\begin{tabular}[c]{@{}l@{}}Red LED\\ (Figure \ref{RedLEDcomparison})\end{tabular}              & 658nm                                                                                  & 657nm                                                                            & 0.15\%                                                                       & \begin{tabular}[c]{@{}c@{}}671.5-637.7= \\ 33.8 nm\end{tabular}                               & \begin{tabular}[c]{@{}c@{}}675.42- 640\\  = 35.42 nm\end{tabular}           & 4.57\%                                                                      \\ \hline
\begin{tabular}[c]{@{}l@{}}Green LED\\ (Figure \ref{GreenLEDcomparison})\end{tabular}            & 553nm                                                                                  & 555nm                                                                            & 0.36\%                                                                       & \begin{tabular}[c]{@{}c@{}}584.62-531.5=\\ 53.12nm\end{tabular}                               & \begin{tabular}[c]{@{}c@{}}581-530.54\\ =50.46nm\end{tabular}               & 5.39\%                                                                      \\ \hline
\begin{tabular}[c]{@{}l@{}}White LED\\ (Figure \ref{WhiteLEDcomparison})\end{tabular}            & \begin{tabular}[c]{@{}c@{}}451nm \\ \&\\  558 nm\end{tabular}                          & \begin{tabular}[c]{@{}c@{}}454nm \\ \& \\ 559 nm\end{tabular}                    & \begin{tabular}[c]{@{}c@{}}0.66\%  \\ \& 0.18\%\\  respectively\end{tabular} & \begin{tabular}[c]{@{}c@{}}474.15-438.4\\ =35.75nm \&\\ 601.85-528.88=\\ 72.97nm\end{tabular} & \begin{tabular}[c]{@{}c@{}}476-439=37nm \\ \& \\ 600-523=77nm\end{tabular}  & \begin{tabular}[c]{@{}c@{}}3.37\% \& \\ 5.32\% \\ respectively\end{tabular} \\ \hline
\end{tabular}
\end{table*}

\section{Obtained Results}

Table \ref{Experimental Data} shows the peak wavelength(s) and line widths obtained for various LEDs and green laser available at market.The outputs match very closely to the ideal ranges for these light sources.In case of green laser,the output peak occurs within the range of it's rating. Figure \ref{allSpectra} show the output graphs obtained during live video streaming using Pycharm platform.

\begin{table*}[]

\centering

\caption{Total Cost Analysis}
\label{Cost table}
\renewcommand{\arraystretch}{1.5}

\begin{tabular}{|c|c|c|}
\hline
\textbf{Component/software} & \textbf{Total cost of our spectrometer} & \textbf{Optical Spectrometer (Red Tide USB650 Fiber Optic Spectrometer)}                             \\ \hline
C270 HD Webcam              & 1900 BDT                                & \cellcolor[HTML]{FFFFFF}{\color[HTML]{191919} }                                                      \\ \cline{1-2}
Housing for webcam          & 800 BDT                                 & \cellcolor[HTML]{FFFFFF}{\color[HTML]{191919} }                                                      \\ \cline{1-2}
Additional costs (DVD)      & 100 BDT                                 & \cellcolor[HTML]{FFFFFF}{\color[HTML]{191919} }                                                      \\ \cline{1-2}
Software (Python)           & Free                                    & \cellcolor[HTML]{FFFFFF}{\color[HTML]{191919} }                                                      \\ \cline{1-2}
Total                       & \textbf{2800 BDT}                       & \multirow{-5}{*}{\cellcolor[HTML]{FFFFFF}{\color[HTML]{191919} \textbf{1,499.99 USD == 176280 BDT}}} \\ \hline
\end{tabular}
\end{table*}

\section{Comparison between our Spectrometer \& Red Tide USB650 Fiber Optic Spectrometer}

Table \ref{Comparison table} shows a detailed comparison between the data obtained from our custom-made low cost spectrometer and from optical spectrometer (Red Tide USB650 Fiber Optic Spectrometer) in laboratory.

Table \ref{Cost table} represents the overall cost analysis of our custom-made spectrometer and an overall comparison of the cost between our spectrometer and Optical Spectrometer  (Red Tide USB650 Fiber  Optic Spectrometer) used in Laboratory.

\section{Conclusion}

In this work, we created a low-cost optical spectrometer with great precision, equivalent to industry standards such as the RED TIDE USB650 Fiber Optic Spectrometer. Using a webcam and a DVD diffraction grating, the spectrometer measured peak wavelengths and line widths of numerous light sources, including LEDs and lasers, with amazing precision. The RED TIDE USB650 was used for calibration, and the device was programmed to automatically adapt for different wavelength ranges, guaranteeing reliable spectral data. Our investigation, which was carried out using the PyCharm platform, revealed that the spectrometer's results differed minimally from those of the professional-grade spectrometer, as seen by the minor proportion of mistakes provided in Table 2. The detailed cost research demonstrated that this spectrometer's design is extremely cost-effective, making it an excellent choice for educational and practical applications. Despite its simplicity and low cost, the spectrometer demonstrated the capacity to properly detect emission wavelengths and spectrum features of unknown sources. This cost-performance balance emphasizes the device's significance, particularly in resource-constrained environments, by providing a feasible option for accurate spectrum analysis.

\bibliographystyle{ieeetr}
\bibliography{references}


\IfFileExists{output.bbl}{
}{}

\end{document}